\documentclass[mathleft,
]{an}
\usepackage{graphicx}
\usepackage{color}
\usepackage{amsmath}
\usepackage{times}
\usepackage{natbib}
\bibpunct{(}{)}{;}{a}{}{,}
\newcommand{\aap}{A\&A}
\newcommand{\aaps}{A\&AS}
\newcommand{\nat}{Nature}

\newcommand{\pasa}{PASA}
\newcommand{\aapr}{A\&A~Rev.}

\newcommand{\angstrom}{\,\mbox{\normalfont\AA}~}

\newcommand\kms{{\rm\,km\,s^{-1}}}

\newcommand\msun{\rm\,M_\odot}
\newcommand\lsun{\rm\,L_\odot}
\newcommand\mlsun{\rm\,M_\odot\,L_\odot^{-1}}
\newcommand\micron{\,\mu{\rm m}}

\begin{document}

\Pagespan{0}{0}
\Yearpublication{2014}%
\Yearsubmission{2014}%
\Month{6}%
\Volume{335}%
\Issue{5}%

\title{Observational Dynamics of Low-Mass Stellar Systems\thanks{Partly based on observations obtained at the European Southern Observatory, Chile (Observing Programmes 078.B-0496(B) and 081.B-0282)}\\
        {\normalsize Doctoral Thesis Award Lecture 2013}
}

\author{Matthias~J.~Frank\inst{1}\fnmsep\thanks{Corresponding author:
  \email{mfrank@lsw.uni-heidelberg.de}}}
\titlerunning{Observational Dynamics of Low-Mass Stellar Systems}

\authorrunning{M.~J.~Frank}
\institute{
Landessternwarte, Zentrum f\"ur Astronomie der Universit\"at
Heidelberg, K\"onigsstuhl 12, D-69117 Heidelberg, Germany}

\received{20 Jan 2014}
\accepted{11 Nov 2005}
\publonline{later}

\keywords{galaxies: compact -- galaxies: dwarf -- galaxies: evolution -- globular clusters: general -- galaxies: kinematics and dynamics}

\abstract{The last fifteen years have seen the discovery of new types of low-mass stellar systems that bridge the gap between the once well-separated regimes of galaxies and of star clusters. Whether such objects are considered galaxies depends also on the definition of `galaxy', and several possible criteria are based on their internal dynamics (e.g. the common concept that galaxies contain dark matter). Moreover, studying the internal dynamics of low-mass stellar systems may also help understand their origin and evolutionary history.
The focus of this paper is on two classes of stellar systems at the interface between star clusters and dwarf galaxies: ultra-compact dwarf galaxies (UCDs) and diffuse Galactic globular clusters (GCs). A review of our current knowledge on the properties of UCDs is provided and dynamical considerations applying to diffuse GCs are introduced. In the following, recent observational results on the internal dynamics of individual UCDs and diffuse Galactic globular clusters are presented.
}

\maketitle

\section{Introduction}
\label{sec:intro}
For many decades the known stellar systems in the universe could be relatively easily classified into either galaxies or star clusters, based on their morphology (late-type galaxies, i.e. spirals and irregulars) and their physical extent in the case of an early-type, i.e. spheroidal or elliptical, morphology. Galaxies, cosmological structures, which formed in dark matter halos and grew via mergers with and accretion of other galaxies, typically have half-light radii ranging from hundreds of pc to tens of kpc. Star clusters, which formed from molecular clouds inside these galaxies, have typical half-light radii of a few to a few tens of pc. This distinction is reflected, for example, in \citeauthor{1994ESOC...49....3T}'s \citeyearpar{1994ESOC...49....3T} widely adopted definition of dwarf galaxies, to be  `fainter than M$_B\!\le\!-16$\,mag and \emph{more extended than globular clusters}'.

However, in recent years several discoveries blurred the boundary between the regime of galaxies and star clusters:
\begin{itemize}
\item Ultra-compact dwarf galaxies (UCDs; Section~\ref{sec:UCDs}) began to fill the parameter space (in size, mass and luminosity) between classical globular clusters on the one side and compact and dwarf ellipticals on the other side.
\item Ultra-faint and extended dwarf spheroidal satellites were discovered in wide-field photometric surveys around our Galaxy and M\,31  \citep[e.g.][]{2004ApJ...612L.121Z,2005AJ....129.2692W,2006ApJ...647L.111B,2006MNRAS.371.1983M} and, apart from containing large amounts of dark matter, resemble diffuse globular clusters. 
\item More and more extended star clusters were detected around nearby galaxies \citep[e.g.][]{2000AJ....120.2938L,2002AJ....124.1410B,2005MNRAS.360.1007H,2009AJ....137.4361D}. With absolute $V$-band magnitudes of $-8\!\le\!\mathrm{M}_V\!\la\!-4$\,mag and projected half-light radii of $5\!\la\!r_{h}\!\la\!25$\,pc they populate the `extended side' of the star cluster distribution \citep[see e.g. fig.~8 of][]{2011AJ....142..199B}.
\item Finally, if also late-type morphologies are considered, tidal dwarf galaxies, which formed in gas-rich galaxy mergers \citep[e.g.][]{1992Natur.360..715B,1998A&A...333..813D} and are thought to be mostly dark matter-free present another challenge to the picture of galaxies residing in dark matter halos on the one side and star clusters on the other side.
\end{itemize} 

As a result, astronomical conferences are being dedicated to the stellar systems at the interface between star clusters and galaxies \citep[e.g.][]{2011Msngr.144...44M} and the question `What is a galaxy?' is being actively discussed in the literature \citep[e.g.][]{2011PASA...28...77F,2012AJ....144...76W}. 

The consensus is that a galaxy is at least gravitationally bound and contains stars, and several additional requirements have been proposed as \emph{sufficient} defining criteria. \citet{2012AJ....144...76W} proposed that (\textsc{i}) 
galaxies are objects, whose properties cannot be explained by a combination of baryons and Newton's laws of gravity, and as a more observationally accessible proxy criterion, that (\textsc{ii}) galaxies show a spread in metallicity [Fe/H]. \citet{2011PASA...28...77F} proposed that galaxies (\textsc{iii}) have a two-body relaxation time greater than the Hubble time, that (\textsc{iv}) galaxies have half-light radii $\ge\!100$\,pc, (\textsc{v}) contain complex stellar populations, (\textsc{vi}) contain non-baryonic dark matter, or (\textsc{vii}) host a satellite system.

It is obvious (criteria \textsc{i},\textsc{iii},\textsc{vi}) that the internal dynamics of an object play a prominent role in whether we consider it a star cluster or a galaxy. Beyond the problem of classification, understanding the inner stellar motions of an individual object may shed light on its origins. In this context, this paper focuses on the internal dynamics of two classes of objects in the border regime between galaxies and star clusters: ultra-compact dwarf galaxies (UCDs, Section~\ref{sec:UCDs}) and diffuse globular clusters (GCs) in the outer Galactic halo (Section~\ref{sec:GalacticGCs}). Following an overview of the properties of UCDs in Section~\ref{sec:UCDsproperties}, some recent observational results on individual UCDs are reported: Section~\ref{sec:ucd3} discusses the first spatially resolved kinematics of a UCD, the luminous Fornax cluster UCD3. Section~\ref{sec:coindex} describes an observational test of the hypothesis that the, on average, too high masses of UCDs are caused by an overabundance of low-mass stars. Section~\ref{sec:GalacticGCs} provides some background on Galactic GCs, and their internal dynamics, in particular of extended outer halo clusters. Recent results on the dynamical state of two such extended outer halo GCs, Palomar~4 and Palomar 14 are presented in Section~\ref{sec:massseg}, and a comparison of the dynamical mass and photometric mass of Palomar~4 in the context of testing for Modified Newtonian Dynamics is summarised in Section~\ref{sec:pal4}. Section~\ref{sec:summary} concludes with a brief summary.

\section{Ultra-compact dwarf galaxies (UCDs)}
\label{sec:UCDs}
\subsection{Properties of UCDs}
\label{sec:UCDsproperties}

\subsubsection{Morphological classification and occurrence}
\label{sec:UCDsmorph}
UCDs were first discovered in spectroscopic surveys of the Fornax cluster \citep{1999A&AS..134...75H,2000PASA...17..227D} as a population of compact stellar systems that are more luminous than any Galactic globular cluster, but significantly more compact than compact ellipticals or dwarf ellipticals of comparable luminosity. These objects constituted the prototypes of the morphological class of `ultra-compact dwarf galaxies'\footnote{Alternative names for these objects exist in the literature, such as `Dwarf-Globular transition objects' \citep[DGTOs;][]{2005ApJ...627..203H}, or simply `ultra-compact objects' \citep[UCOs;][]{2002A&A...383..823M} or `compact stellar systems' \citep[CSSs;][]{2009MNRAS.394.1801F}}. \citep[UCDs;][]{2001ApJ...560..201P}, which can be loosely defined as predominantly old \citep[e.g.][]{2011MNRAS.412.1627C}, early-type stellar systems with half-light radii of $15\!\la\!r_h\!\la\!100$\,pc and absolute luminosities of $-13.0\!\la\! M_V\!\la\!-11$\,mag, or alternatively, masses of $2\times10^6\!\ga\!\mathrm{M}\!\ga\!2\times10^8\,\msun$ \citep{2008ApJ...677..276M}.

Following their initial discovery, systematic searches for UCDs were conducted, and UCDs were also discovered in other all-target spectroscopic surveys of galaxy clusters. These studies yielded a larger number of known UCDs in the Fornax cluster \citep{2007A&A...463..119H,2007A&A...463..503M,2009AJ....137..498G} and objects of this type have also been found in other nearby galaxy clusters, such as Virgo \citep{2005ApJ...627..203H,2006AJ....131..312J,2011AJ....142..199B,2013ApJ...775L...6S}, Antlia \citep{2013MNRAS.430.1088C}, Centaurus \citep{2007A&A...472..111M}, Coma \citep{2009MNRAS.397.1816P,2010ApJ...722.1707M,2011ApJ...737...86C}, Hydra I \citep{2007ApJ...668L..35W,2011A&A...531A...4M}, and Perseus \citep{2012MNRAS.422..885P}. More recently objects of this type have also been discovered in less rich environments, around the dominant giant galaxies in nearby loose groups, namely NGC~5128 \citep{2007A&A...469..147R,2010ApJ...712.1191T}, M~104 \citep{2009MNRAS.394L..97H}, NGC~4546 and NGC~3923 \citep{2011MNRAS.414..739N}, NGC~4494 \citep{2011MNRAS.415.3393F}, as well as in the compact groups HCG~22 and HCG~90 \citep{2011A&A...525A..86D} and in the fossil group NGC~1132 \citep{2011ApJ...737L..13M}. 

Up to now, no UCDs are known that exist in isolation, but this is perhaps not surprising: due to their small sizes, most UCDs appear unresolved from the ground and thus evade detection in wide-field imaging surveys. The identification of UCDs therefore has been based on all-target spectroscopic surveys of galaxy groups or clusters (where UCD candidacy of unresolved sources was inferred by proximity to the cluster galaxies on the sky) or on follow-up spectroscopic observations of candidates identified as marginally resolved objects in the surroundings of galaxies imaged by the Hubble Space Telescope (HST). If isolated UCDs exist at all, there is a small chance to discover them by their radial velocity in blind spectroscopic surveys of unresolved sources with colours of late spectral type stars, such as SEGUE \citep{2009AJ....137.4377Y} or the stellar surveys conducted with LAMOST \citep{1998IAUS..179..131C}.

\subsubsection{Formation Scenarios}
\label{sec:UCDsformation}
Several scenarios for the formation and nature of UCDs have been proposed in the literature. UCDs could be the remnant nuclei of tidally stripped galaxies. Based on numerical simulations of this `threshing' scenario, \citet{2001ApJ...552L.105B} and \citet{2003MNRAS.344..399B} showed that the tidal stripping of nucleated dwarf ellipticals, as well as of nucleated low-surface brightness spiral galaxies, on eccentric orbits and in close passages (pericenter distances of 5--70\,kpc) of the Fornax cluster centre, can result in naked nuclei that resemble observed UCDs. Qualitatively similar results were obtained by \citet{1994ApJ...431..634B}, who studied the formation of globular clusters from nucleated dwarf galaxies. In particular, depending on the eccentricity and pericenter distance of the orbit and the dark matter distribution of the original nucleated galaxy, the resulting UCDs can also retain low-surface brightness envelopes, similar to the extended halos observed around some UCDs \citep{2008AJ....136..461E}. In the threshing scenario, UCDs would resemble lower-mass versions of M32-like compact ellipticals, the majority of which are believed to have formed via tidal stripping (e.g. \citealt{2009Sci...326.1379C}; but see \citealt{2013MNRAS.430.1956H}, for an apparent counterexample).

UCDs could also be massive star clusters. \citet{2002A&A...383..823M} proposed that UCDs represent the high-luminosity tail of the (intra-cluster) GC population. \citet{2002MNRAS.330..642F} proposed that UCDs could be the result of mergers of stellar superclusters \citep[i.e. aggregates of many young star clusters;][]{1998MNRAS.300..200K} that formed in gas-rich galaxy-galaxy mergers. The massive star cluster W3 in the merger remnant galaxy NGC~7252 may be an example for the latter scenario \citep{2005MNRAS.359..223F}. With a mass of $\sim\!10^8\,\msun$ and a half-light radius of 18\,pc it resembles typical UCDs, except for its young age of only $\sim\!0.5$\,Gyr.

\subsubsection{Spatial, velocity and luminosity distribution}
\label{sec:UCDsdistrib}
Beyond the search for UCDs and the confirmation of UCD candidates, studies of these objects, mainly in the Fornax and Virgo clusters, have focused on collecting information on their bulk properties and discussing these mainly in the context of the `threshed nuclei' and `star clusters' scenarios.
In Fornax, the spatial and kinematic distribution of UCDs is distinct from that of nucleated and non-nucleated dwarf galaxies and the GC systems of the central cluster galaxies \citep[e.g.][]{2003Natur.423..519D,2004PASA...21..375D,2004A&A...418..445M,2012A&A...537A...3M,2009AJ....137..498G}. Since simulations of the threshing scenario require highly eccentric orbits to form UCDs and since the remnant nuclei will remain on these orbits, \citet{2004PASA...21..375D} argued that the more concentrated distribution of UCDs compared to nucleated dwarfs disfavours these as the UCD progenitors, and that UCDs therefore may represent the progeny of a novel class of galaxies. A potential caveat of this argument is that there is a selection bias. The nucleated dwarfs observed at the present are those that were \emph{not} tidally stripped, and therefore these could be on orbits less susceptible to threshing, or could have fallen into the cluster at a later time. \citet{2009AJ....137..498G} showed that the spatial distribution of UCDs in Fornax is compatible with that of the intra-cluster GCs, i.e. GCs not associated with one of the central cluster galaxies. Also in other galaxy clusters, the specific frequency of UCDs is found to be very similar to that of GCs, consistent with UCDs being the bright end of the GC luminosity distribution \citep{2012A&A...537A...3M}. 

\subsubsection{Structure of UCDs}
\label{sec:UCDsstruc}
At a given luminosity, UCDs are more extended than the nuclei of galaxies \citep[e.g.][]{2005ApJ...623L.105D,2008AJ....136..461E}, which has been put forward as an argument against the threshing model. However, also in the threshing simulations, a certain degree of expansion of the remnant nuclei is seen \citep[see e.g. fig. 4 of][]{2003MNRAS.344..399B}. Quantifying this effect, \citet{2013MNRAS.433.1997P} confirmed that for suitable orbits, the stripping of nucleated dwarf galaxies can produce objects with the full range of sizes observed in UCDs. In contrast to GCs, whose sizes and luminosities are uncorrelated, and in common with galaxies, UCDs may show a trend of increasing size with increasing luminosity \citep[][]{2006AJ....131.2442M,2008AJ....136..461E}. \citet{2011AJ....142..199B}, who include also extended but faint clusters (half-light radii of $5\!\la r_h\!\la25$\,pc and luminosities of $M_V\!\la\!-8$\,mag) in their study, argue that such a size-luminosity relation does not exist and that extended clusters and UCDs show a continuous distribution in luminosity that parallels that of concentrated GCs. However, they also note that there is a very sparsely populated prominent gap in luminosity ($-8<M_V<-11$\,mag) between faint extended clusters and UCDs (see their fig.~8), although this may be a consequence of the target selection in spectroscopic surveys of UCDs and star clusters. Nevertheless, this suggests that caution is advised regarding the existence of a size-luminosity relation.

\subsubsection{Ages and metallicities of UCDs}
\label{sec:UCDspop}
Stellar population studies \citep[e.g.][]{2004A&A...416..467M,2006AJ....131.2442M,2007A&A...463..503M,2007AJ....133.1722E,2009MNRAS.394L..97H, 2010ApJ...712.1191T,2011MNRAS.412.1627C} showed that UCDs have generally sub-solar metallicities of $-2\!\la\!\mathrm{[Fe/H]}\!\la\!0$\,dex and are predominantly old ($\ga\!9$\,Gyr), although there are a few younger UCDs with ages of $0.5$ to several Gyr, such as the already mentioned UCD-like young massive cluster W3 in NGC~7252. Both, the metallicities and the ages derived in these studies, are luminosity-weighted, single stellar population equivalent quantities, and to date the observationally challenging question whether UCDs contain multiple stellar population has not been addressed.

Like the GC systems of many galaxies, also the GC populations of the Virgo and Fornax clusters show a bimodality in colour \citep{2006ApJ...653..193M,2010ApJ...710.1672M}. These bimodalities are known to reflect an actual bimodality in the metallicity distribution in many systems, although in some cases they may also result from the non-linear dependence of colour on metallicity \citep[e.g.][and references therein]{2012A&A...539A..54C}. In terms of colours, UCDs in the Virgo cluster coincide with the blue peak of the cluster's GC distribution and UCDs in the Fornax cluster, being significantly redder, coincide with the red peak of the GC distribution. The same correlations hold also when spectroscopic metallicity, instead of colour, is considered and could indicate a fundamental difference in the origin of UCDs in the Virgo and Fornax clusters \citep{2006AJ....131.2442M}. \citet{2010ApJ...724L..64P} compared the stellar population properties of nuclei of 34 Virgo dwarf ellipticals (dE) and ten Virgo UCDs. In agreement with previous studies of the Virgo cluster, they found that dE nuclei are on average more metal-rich than UCDs. However, they showed that in the high-density regions of the cluster, where most of the UCDs reside, there is no difference in metallicity between nuclei and UCDs. Hence, they argued that the difference in metallicity in the cluster-wide sample can be explained by dEs being able to retain retain gas for a longer time and therefore form younger and more chemically enriched generations of stars in the lower-density regions of the cluster. UCDs in Virgo could therefore descend from dE nuclei in which the star formation, due to the higher environmental density, was shut down at earlier epochs. On the other hand, \citet{2006AJ....131.2442M} based on their finding that nuclei in the Fornax cluster are significantly more metal-poor and potentially older than the UCDs in this cluster, argue that UCDs in Fornax may be the successors of merged stellar superclusters produced in violent galaxy-galaxy mergers.

Galactic GCs \citep[e.g.][]{2011MNRAS.412.2199T}, as well as massive elliptical galaxies \citep{1998ApJS..116....1T,2000MNRAS.315..184K}, are known to have stellar populations enhanced in $\alpha$-elements compared to the solar ratio, indicating a rapid star formation time scale. Dwarf ellipticals and also the nuclei of nucleated dEs in the majority have sub-solar to solar [$\alpha$/Fe] ratios and therefore are thought to have experienced prolonged star formation \citep[e.g.][]{2003AJ....126.1794G,2007AJ....133.1722E}. \citet{2007AJ....133.1722E} find that the majority of a sample UCDs in Virgo shows super-solar [$\alpha$/Fe] ratios, compatible with a star-cluster origin of UCDs. \citet{2009MNRAS.394.1801F} find a more diverse picture for a small sample of Fornax UCDs, with at least one UCD, UCD3 (see Section~\ref{sec:ucd3}) having at most solar [$\alpha$/Fe]. UCDs in NGC~5128 (Centaurus A) appear significantly less enhanced in $\alpha$-elements than the Virgo and Fornax samples \citep{2010ApJ...712.1191T} and therefore more closely resemble dE nuclei.

\subsubsection{`A mixed bag of objects' \citep{2009gcgg.book...51H}}
\label{sec:UCDsconsensus}
In summary, the ensemble properties of UCDs thus far have not provided clear evidence for or against UCDs being either star clusters or threshed nuclei. Consequently, a consensus seems to emerge in the literature, that the morphological class of UCDs comprises objects from different evolutionary tracks \citep[e.g][]{2009gcgg.book...51H,2011A&A...525A..86D,2011MNRAS.414..739N,2011AJ....142..199B}, even  though the fraction of threshed nuclei among UCDs may be very low \citep{2012A&A...537A...3M}.

\subsubsection{High mass to light ratios}
\label{sec:UCDshighML} 
One of the most intriguing properties of UCDs are their high dynamical mass to light ratios (M/L) that are on average $\sim$30-50 per cent higher than the predictions of canonical stellar population models \citep[e.g.][]{2009A&A...500..785K}. While there are UCDs for which the dynamical and stellar population M/L largely agree with each other \citep[e.g.][]{2011MNRAS.412.1627C}, there are also extreme cases, for which the dynamical M/L is up to $\sim3$ times higher than expected from the stellar population synthesis \citep[e.g.][]{2005ApJ...627..203H,2008ApJ...677..276M,2010ApJ...712.1191T}. A recent overview on the dynamical M/L of UCDs was provided by \citet{2013A&A...558A..14M}.

Regardless of the origin of UCDs, their high M/L are not readily explained. If UCDs are threshed nuclei, they would reside in the centre of the (remnant) dark matter halo of the original galaxy. However, in order to affect the stellar kinematics of such compact objects, dark matter halos with extremely high central densities would be required, two orders of magnitude higher than the central dark matter densities of the, presumably cored, dark matter halos of dwarf spheroidals \citep[e.g.][]{2007ApJ...663..948G,2010MNRAS.403.1054D}. This holds also, assuming realistic halo masses, for the central dark matter densities in the cuspy \citet{1997ApJ...490..493N} or \citet{1999MNRAS.310.1147M} halos \citep{2009ApJ...691..946M}. A possible way to alleviate this problem could be the enhancement of the central dark matter concentration during the formation of the progenitor nuclei via the infall of gas \citep{2008MNRAS.385.2136G,2008MNRAS.391..942B}.

Another possibility in the tidal stripping scenario is that the M/L$_V$ of the remnant UCDs are enhanced due to the presence of unbound stars around the UCDs. In this case, the integrated-light velocity dispersion would no longer be that of a system in virial equilibrium and the inferred dynamical mass would be systematically too high. Based on $N$-body simulations \citet{2006MNRAS.367.1577F} showed that this effect can enhance the measured M/L$_V$ of UCDs with masses of $10^7\msun$ by a factor of up to 10, if the line-of-sight coincides with the UCD's trajectory. The measured M/L$_V$ is not affected, if the line-of-sight is perpendicular to the orbital plane, or if the UCD is compact and massive (M$\sim10^8\msun$). Assuming that the orbital planes of UCDs are randomly oriented, it seems difficult to explain that \emph{most} UCDs with M$\!\ga\!10^7\msun$ show an enhanced M/L$_V$. Moreover, unbound stars would eventually escape into the host halo, and as the majority of UCDs are old ($\ga\!9$\,Gyr), it is unlikely that they were only threshed within the last few Gyr.

UCDs could also harbour massive black holes, either because a black hole and a nucleus of comparable mass coexisted in the progenitor galaxy, or because UCDs are in fact `hyper-compact stellar systems' bound to recoiling super-massive black holes that were ejected from the cores of giant galaxies \citep{2009ApJ...699.1690M}. The black hole masses expected in these scenarios, and required to explain the high M/L of UCDs, could substantially exceed those of intermediate-mass black holes that may \citep[e.g.][]{2002ApJ...578L..41G,2002AJ....124.3270G,2008ApJ...676.1008N,2011A&A...533A..36L} or may not \citep[e.g.][]{2003ApJ...589L..25B,2010ApJ...710.1032A,2010ApJ...710.1063V,2013ApJ...769..107L} be present in GCs. \citet{2013A&A...558A..14M} showed that for massive UCDs, black hole masses of $\sim10-15$\,per cent of the UCD mass would explain the difference between dynamical mass and the mass obtained from the luminosity and a population synthesis mass to light ratio. Since this fraction lies well within the range spanned by black hole masses in nuclear clusters \citep{2009MNRAS.397.2148G}, this may support a threshing origin of massive UCDs.   

These three scenarios, i.e. the presence of dark matter, of tidal disturbance, or of a massive black hole, can in principle be tested by obtaining \emph{spatially resolved} kinematics of UCDs. Due to their compact sizes, low velocity dispersions, large distances and therefore faint magnitudes, however, this remains observationally challenging even with current 8\,m class telescopes, because it requires high spatial and spectral resolution. A first study of the spatially resolved kinematics of an UCD was presented by \citet{2011MNRAS.414L..70F} and will be described in Section~\ref{sec:ucd3}. 

An alternative scenario to such extra, dark constituents of the UCDs' masses and to recent tidal interactions is that the stellar populations of UCDs themselves are not `normal', showing an overabundance of underluminous stellar objects. This has been proposed in the form of an overabundance of either stellar remnants \citep[top-heavy mass function;][]{2008MNRAS.386..864D,2009MNRAS.394.1529D,2010MNRAS.403.1054D} or of low-mass stars \citep[bottom-heavy mass function;][]{2008ApJ...677..276M}. \citet{2012ApJ...747...72D} find evidence in favour of the former, top-heavy mass function scenario in the form of a higher abundance of X-ray sources in massive UCDs compared to the expectation from a standard \citet{2001MNRAS.322..231K} initial mass function and an assumed size-luminosity relation for UCDs (cf. Section~\ref{sec:UCDsstruc}). An observational test of the latter, bottom-heavy mass function scenario is presented in Section~\ref{sec:coindex}.

\begin{figure*}
\centering
\includegraphics[width=0.65\textwidth]{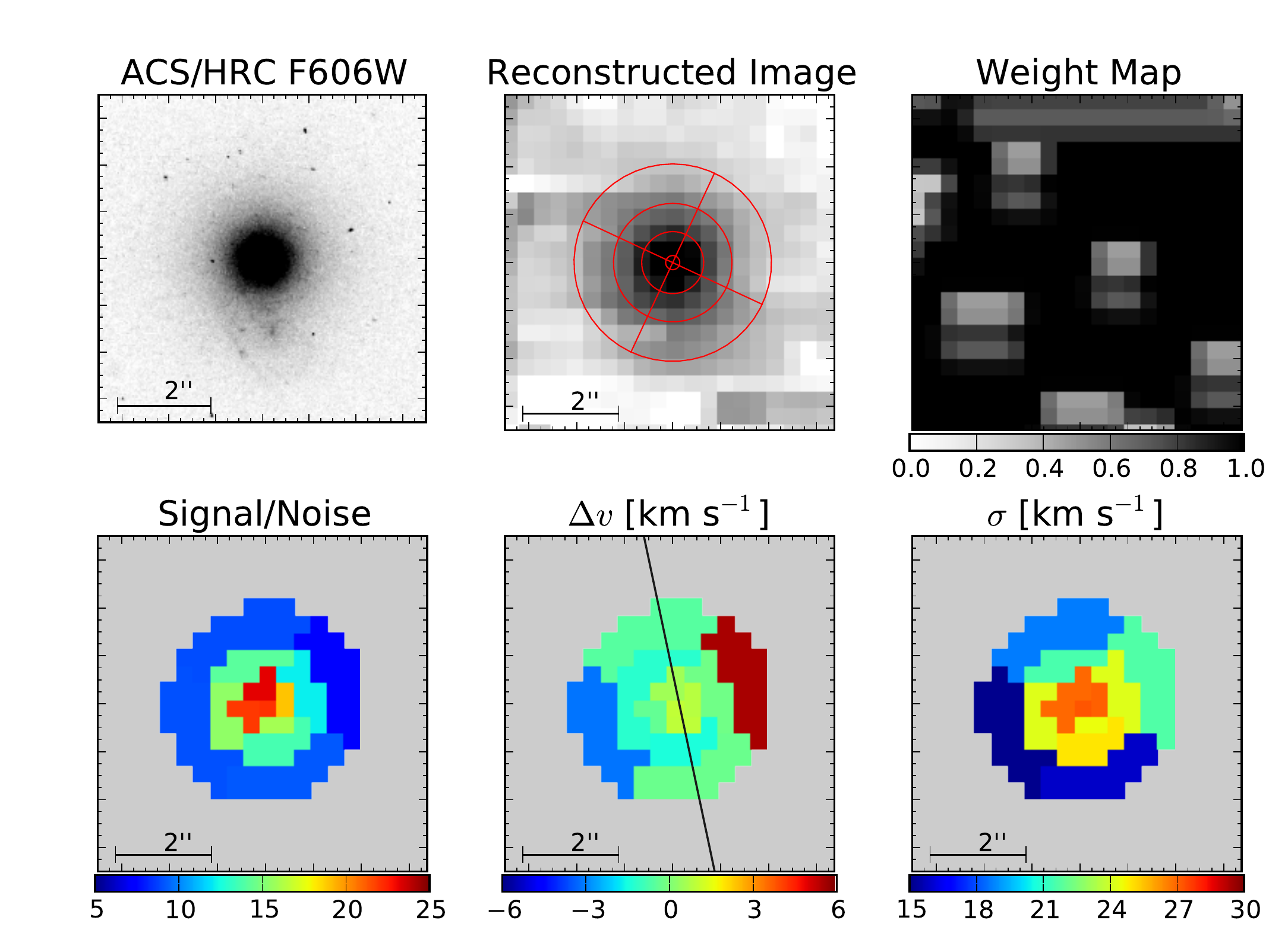}
\caption{The upper row shows an archival HST image (obtained with the ACS/HRC in broad $V$ filter F606W) of the UCD, the image reconstructed from the merged data cube and a weight map, which illustrates the effect of bad spaxels in the input data cubes. The polar grid overlaid on the reconstructed image indicates the subdivisions used for the spatial binning. The position angle of the quadrant subdivisions was chosen such that the rotation of the UCD is visible. The bottom row shows the mean signal to noise ratio in each spatial bin per spectral bin (of 11\,km~s$^{-1}$ or $\sim0.2$\,\AA), the recovered velocity field with a line indicating the axis of rotation and the velocity dispersion map. The scale of the images is indicated by the 2\,arcsec scale bar at the lower left, 1\,arcsec corresponds to $\sim 92$\,pc at the distance of the Fornax cluster.}
\label{ucd3:fig:maps}
\end{figure*}

\subsection{Spatially resolved kinematics of an UCD}
\label{sec:ucd3}
Integral-field spectroscopy has become a well established technique to study the internal kinematics of unresolved stellar systems and has been employed in large surveys of massive galaxies \citep[e.g.][]{2002MNRAS.329..513D,2011MNRAS.413..813C,2012A&A...538A...8S}. However, resolving the kinematics of objects with angular half-light radii on the order of 1\,arcsec, takes seeing-limited optical integral field spectroscopy to its limits. On the other hand, present-day adaptive optics-aided, near infra-red integral field units (IFUs; e.g. Gemini/NIFS; VLT/SINFONI), provide only a moderate spectral resolution ($\sigma_\mathrm{ins}\ga32$\,km~s$^{-1}$), limiting their ability to accurately sample the velocity distribution of typical UCDs with velocity dispersions of $\sim 25$\,km~s$^{-1}$.

To date, only one UCD has been studied using integral-field spectroscopy \citep{2011MNRAS.414L..70F}. The target of this study, UCD3, is one of the defining objects of the class of UCDs \citep{1999A&AS..134...75H}. Still, UCD3 is not an average UCD. With $M_{V}=-13.55$\,mag, it is the brightest known UCD in the Fornax cluster, separated by $\sim0.8$\,mag from the rest of the cluster's UCD population. It is unusually red, with a spectroscopic metallicity of [Fe/H]$\sim-0.2$\,dex \citep{2008MNRAS.390..906C}. Most importantly, it has a large half-light radius of $r_\mathrm{h}=87$\,pc, corresponding to $\sim\!0.95$\,arcsec at a Fornax cluster distance of 18.97\,Mpc \citep{2001ApJ...553...47F}, and is one of the few UCDs where an extended, low surface-brightness envelope has been detected \citep{2008AJ....136..461E}, making it possible to resolve the UCD from the ground with seeing-limited instruments.

\subsubsection{Observations and analysis}
\label{sec:ucd3:observations}
UCD3 was observed in ESO programme 078.B-0496 under excellent seeing with the ARGUS IFU \citep{2003Msngr.113...15K} of the Flames/GIRAFFE \citep{2002Msngr.110....1P} spectrograph mounted on UT2 at the VLT. ARGUS consists of an array of $14\times22$ lenslets. In its `1:1' mode the spatial sampling is $0.52\times0.52$\,arcsec per lenslet (or `spaxel'), which yields a field of view of $11.5\times7.3$\,arcsec. In the LR04 grism, the instrument provides a spectral resolution of R$\!\sim\!9600$ (corresponding to $\sigma_\mathrm{ins}\sim\!13$\,km~s$^{-1}$), over the range $5015\!-\!5831$~\angstrom. Out of eight $2775$ second exposures, the six with the best seeing (measured by by forward-modelling the IFU data from the HST image shown in the upper left panel of Fig.~\ref{ucd3:fig:maps}) were used for the further analysis. 

These data (with a seeing of $0.50\le$ PSF FWHM $\le0.67$\,arcsec, and a mean of $0.60$\,arcsec) were reduced using the \textsc{GIRBLDRS} package \citep{2000SPIE.4008..467B}, sky-subtracted and corrected for atmospheric dispersion. A merged data cube was created by combining the individual data cubes while masking bad spaxels and spectra contaminated by light from neighbouring calibration lamp fibres (illustrated by the weight map shown in Fig.~\ref{ucd3:fig:maps}). Individual spaxels were binned following the geometric binning scheme super-imposed on the reconstructed image in Fig.~\ref{ucd3:fig:maps}, in order to achieve a sufficient signal to noise ratio in the outer parts of the object. 

The velocity $v$ and velocity dispersion $\sigma$ were measured from each binned spectrum using the penalised pixel-fitting (\textsc{ppxf}) code of \citet{2004PASP..116..138C}. Spectra of late-type stars from the UVES Paranal Observatory Project \citep[UVESPOP,][]{2003Msngr.114...10B} library, degraded to the resolution of ARGUS, served as templates in the pixel-fitting.

\subsubsection{Results}
\label{sec:ucd3:results}

\begin{figure}
\centering
\includegraphics[width=0.9\columnwidth]{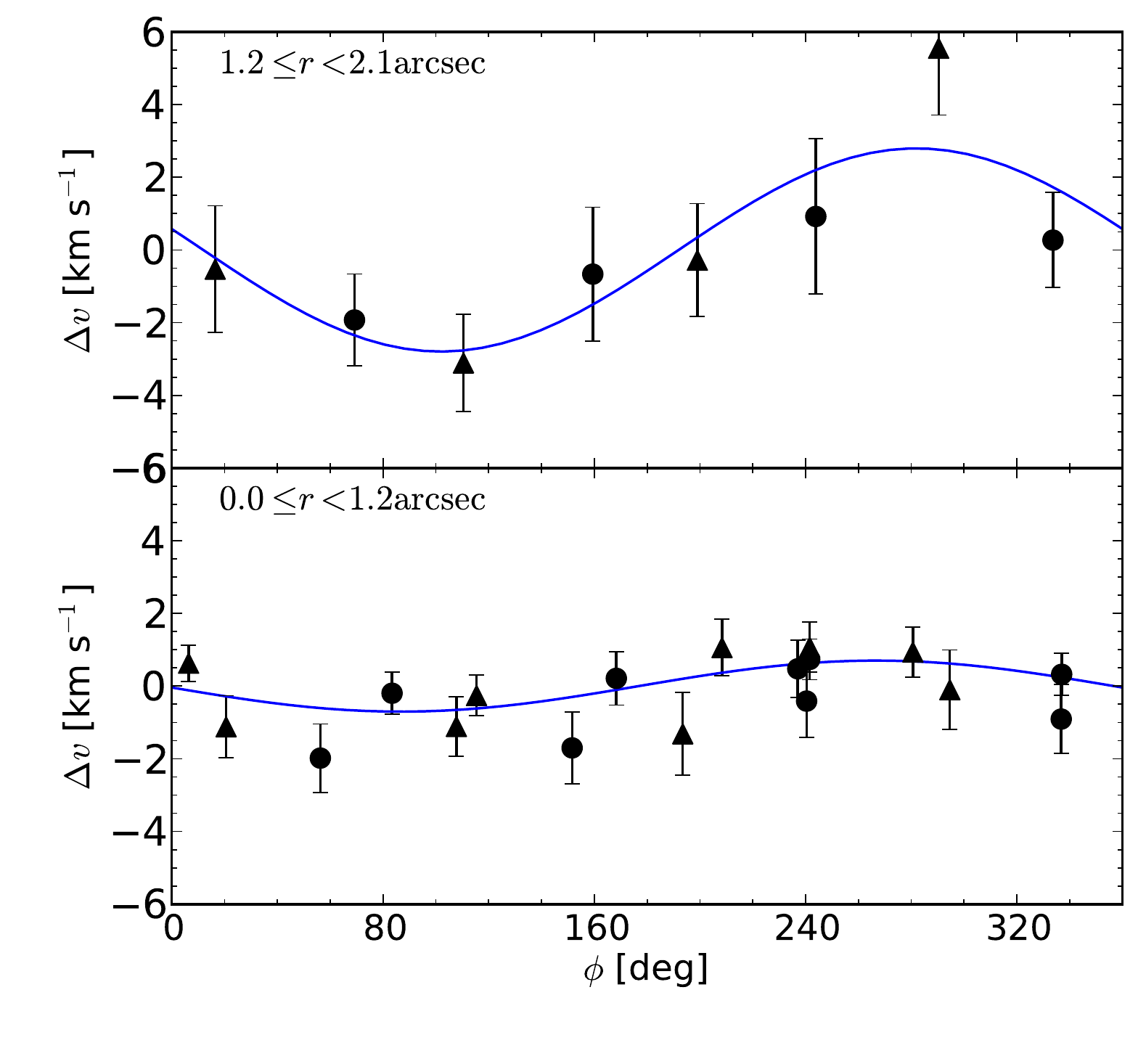}
\caption{The signature of rotation in UCD3. The top panel shows the measured velocity versus the luminosity-weighted azimuth $\phi$ of each bin in the outer annulus (r$\ge1.2$\,arcsec) of the spatial binning shown in Fig.~\ref{ucd3:fig:maps} (triangles) and the binning orientation with quadrants rotated by $45$ degrees. The blue curve represents the best-fitting sinusoid to these data. The bottom panel shows the same for the inner annuli (r$<1.2$\,arcsec).}
\label{ucd3:figrotation}
\end{figure}

The velocity field (lower middle panel of Fig.~\ref{ucd3:fig:maps}) of the UCD shows the signature of weak rotation. The orientation of the rotational axis was recovered by varying the position angle of the quadrants in the spatial binning. Fig.~\ref{ucd3:figrotation} shows the radial velocity as a function of azimuth, in the outer annulus (r$\ge1.2$\,arcsec; upper left panel), as well as in the inner region (r$<1.2$\,arcsec). The ratio of the maximum measured rotational velocity to the central velocity dispersion, $v_\mathrm{rot,max}/\sigma_0\approx0.1$, is lower than observed in massive Galactic GCs \citep[e.g.][]{2013ApJ...772...67B}. However, due to the unknown inclination of the rotational axis and due to the fact that the maximum of the rotation curve velocity need not be reached at r$\sim1.5$\,arcsec, $v_\mathrm{rot,max}=2.8\pm0.7$\,km~s$^{-1}$ represents a lower limit on the true rotational velocity.

\begin{figure}
\centering
\includegraphics[width=0.96\columnwidth]{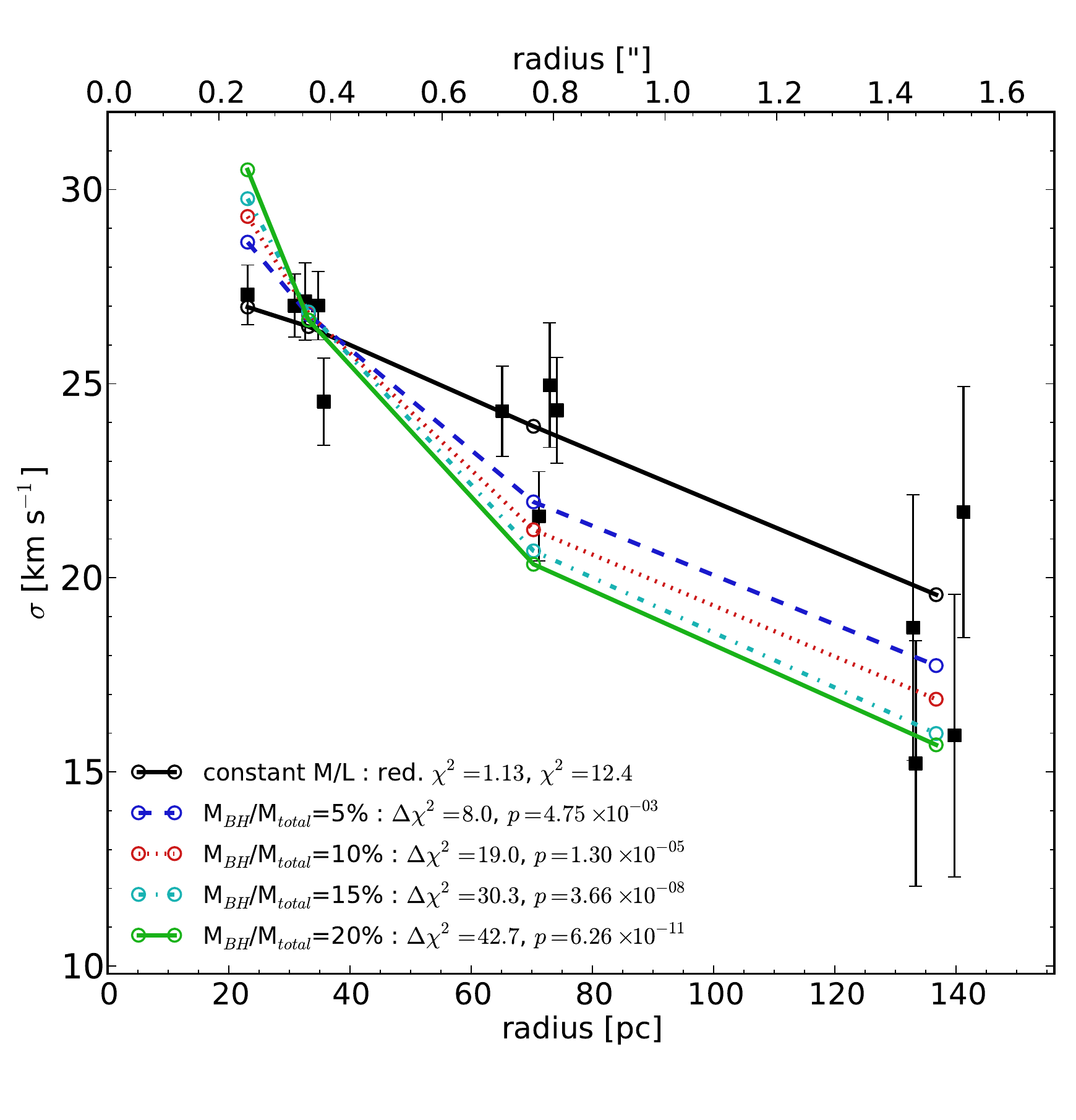}
\caption{The observed velocity dispersion profile (black squares) of UCD3, along with models. The solid black curve shows a constant M/L model, the blue dashed, red dotted, cyan dash-dotted and green solid curves represent the best-fitting models that in addition to the stellar component include a black hole mass of 5, 10, 15 and 20 per cent of the UCD's total mass, respect. The difference in $\chi^2$ with respect to this best-fitting model, $\Delta\chi^2$, along with the resulting probability for the model to represent the data is reported in the legend for each model. The best-fitting mass to light ratios are $2.72\pm0.07$, $2.30\pm0.06$, $1.96\pm0.05$, and $1.78\pm0.05\mlsun$ for M$_\mathrm{BH}$/M$_\mathrm{total}$ of 5, 10, 15, and 20 per cent, respectively. The radial velocity dispersion profile predicted by models including a massive black hole are steeper than the observed profile.}
\label{ucd3:figdispprofile}
\end{figure}

The velocity dispersion profile of UCD3, i.e. the measured velocity dispersion versus the luminosity-weighted radial coordinate of each spatial bin, is shown as black squares in Fig.~\ref{ucd3:figdispprofile}. Except for the innermost radius, where there is only one spaxel, the four measurements at a given radius correspond to the four quadrants in a given annulus. The velocity dispersion decreases continuously with increasing radius. There is no sign of a significantly disturbed profile, or for an increase of the dispersion at large radii, which could have been expected if the unbound stars were present due to a recent tidal interaction of the UCD. 

Along with the observed velocity dispersion profile, Fig.~\ref{ucd3:figdispprofile} shows a model assuming that mass follows light. Open black circles correspond to the model prediction in the radial annuli of our binning, the solid black curve simply connects these model points for clarity. The dynamical models are based on the best-fitting two-component luminosity distribution determined by \citet{2007AJ....133.1722E}, a Fornax cluster distance of 18.97\,Mpc \citep{2001ApJ...553...47F} and assuming an isotropic velocity distribution and spherical symmetry. They were constructed using the code described in \citet{2007A&A...463..119H}, which takes into account the convolution with the seeing PSF and predicts the full line-of-sight velocity distribution in each of the radial annuli. 

The best-fitting mass (or equivalently the best-fitting M/L) was obtained by scaling the dynamical model to the data in a minimum-$\chi^2$ sense. The simplest possible model (black curve), i.e. assuming that mass follows light (or equivalently, a constant M/L) provides an excellent fit to the shape of the observed dispersion profile. It yields a best-fitting mass of $\mathrm{M}=8.2\pm0.2\times10^{7}$\,M$\odot$, which corresponds to a mass to light ratio of $\mathrm{M}/\mathrm{L}_{V,\mathrm{dyn}}=3.6\pm0.1\mlsun$. This M/L is in excellent agreement with the stellar population mass to light ratio of M/L$_{V}=3.7\pm0.2\mlsun$ based on the most recent determination of the UCD's stellar population age and metallicity by \citet{2011MNRAS.412.1627C}. The mass is also in good agreement with the dynamical mass obtained from a single integrated spectrum of the object \citep{2007A&A...463..119H}.

\label{ucd3:sec:bhmodels}

As discussed in Section~\ref{sec:UCDshighML}, the presence of a massive black holes may explain the, on average, high dynamical masses of UCDs. A family of mass models which, in addition to the constant M/L stellar component, includes massive black holes is also shown in Fig.~\ref{ucd3:figdispprofile}. The blue dashed, red dotted, cyan dash-dotted and green solid curves represent the best-fitting models with a black hole mass of 5, 10, 15 and 20 per cent of the UCD's total mass, respectively. It is evident, that the black hole models predict too steep a dispersion profile compared to the observations. Already the model with a black hole of 5 per cent of the UCD's mass, which yields a best-fitting stellar M/L$_{V,\mathrm{dyn}}$ of $2.7\pm0.1\msun$, is excluded at the $>3\sigma$ confidence level.

Similar constraints are obtained for models including dark matter (not shown). Including a dark matter component with a  dark matter distribution mimicking the final evolutionary stage in the threshing simulations of \citet{2008MNRAS.385.2136G}, yields e.g. that a dark matter fraction of 20 per cent within a three-dimensional radius of 200\,pc is compatible with the velocity dispersion profile at the $3\sigma$ level. However, this and any models with a higher amount of dark matter are clearly shallower than both the observed dispersion profile and the mass-follows-light model.

In summary, the brightest known UCD in the Fornax cluster, UCD3, shows weak rotation and a radial dispersion profile in excellent agreement with an isotropic velocity distribution and a constant mass to light ratio. The shape of the radial velocity dispersion profile does not show any sign for tidal disturbance. Moreover, it excludes the presence of a black hole with a mass $\ga5$ per cent of the total mass, and of dark matter contributing $\ga20$ per cent of the mass within a three-dimensional radius of 200\,pc. Hence UCD3 does not show any of the attributes that would strongly suggest that the UCD is the remnant nucleus of a stripped galaxy. While this result does not exclude such scenarios, the internal kinematics of UCD3 are fully consistent with it being a massive globular cluster.

\subsection{Test for a bottom-heavy stellar mass function in UCDs}
\label{sec:coindex}
The universality of the initial stellar mass function (IMF) has been subject of a long standing debate \citep[see][for a review]{2010ARA&A..48..339B}. Recently, a number of studies presented indications for a variation of the IMF in massive galaxies. Based on spatially resolved kinematics of a sample of massive elliptical galaxies, \citet{2012Natur.484..485C} found that the mass to light ratio of their stellar component increases with increasing stellar mass and differs from the expectation from population synthesis models based on a \citet{2001MNRAS.322..231K} or \citet{2003PASP..115..763C} IMF. They argued that this discrepancy can be explained with either a bottom-heavy or top-heavy stellar mass function. Similar deviations were also found by comparing gravitational lensing masses to galaxy luminosities \citep[e.g.][]{2010MNRAS.402L..67G,2011MNRAS.417.3000S}. In addition to these indirect indications for a variation of the IMF, \citet{2010Natur.468..940V,2011ApJ...735L..13V} presented more direct evidence for such a variation. By measuring the strength of gravity-sensitive features in the integrated spectra of massive early-type galaxies they derived a significant overabundance of low-mass stars in the cores of these galaxies, indicating a bottom-heavy stellar mass function. 

\citet{2008ApJ...677..276M} proposed a similar test for a bottom-heavy stellar mass function in UCDs. Assuming that the high dynamical masses of UCDs are entirely due to a bottom-heavy stellar mass function, they predicted that the UCDs with the highest M/L ratios should show a significantly weaker absorption in the gravity-sensitive $2.3\micron$ CO band. This was followed up observationally (Frank et al., in prep.) using low-resolution $K$-band spectroscopy aimed at measuring the strength of the CO absorption in UCDs with particularly high M/L, as well as in comparison objects with lower M/L.

\subsection{Observations and analysis}
\label{sec:coindex:obs}
The spectroscopic data were obtained in ESO programme 081.B-0282 using the short-wavelength arm of the ISAAC spectrograph \citep{1998Msngr..94....7M} with the low resolution grating and 1\,arcsec slit, which yields a resolving power of $R=450$ in the $K$-band ($\lambda \sim$1.82 - 2.5$\mu m$). The primary targets, the Virgo cluster UCDs S928 and S999, which have a dynamical M/L exceeding the stellar population M/L by a factor of $\Psi=\mathrm{M}_\mathrm{dyn}/\mathrm{M}_\mathrm{pop}=$2.3 and 4.4, respectively \citep{2013A&A...558A..14M}, were observed for 8.2\,hours in total. For comparison,  four massive GCs/UCDs in NGC~5128, with lower M/L ($\Psi=0.6-1.6$) were observed for 0.3--0.4\,hours each. 

The nodded data frames were combined, flat-fielded and wavelength-calibrated using the ISAAC instrument pipeline. The flux calibration and telluric correction was derived from telluric standards taken with the science data. 

Based on simulations of the data, it was found that standard methods of spectrum extraction \citep[i.e. the summation over an aperture, or optimal extraction; ][]{1986PASP...98..609H} yielded unreliable results in the low signal to noise regime of the S928 and S999 observations. Thus, instead of extracting one-dimensional spectra, the calibrated and stacked two-dimensional spectral images were first summed along the wavelength direction over the feature and continuum passbands defining the photometric CO index (see below). Counts were then extracted from these summed `spatial' stripes. This yielded unbiased results, but at the cost of substantially larger measurement uncertainties. 

Historically based on two narrow band filters \citep{1975ApJ...200L.123F}, the CO index can be defined as, 
\begin{equation}
\label{coindex:eq:COindex}
\mathrm{CO} = -2.5 {\rm log} \frac{\int^{2.4\mu{\rm m}}_{2.32\mu{\rm m}} f_\lambda d\lambda}{\int^{2.255\mu{\rm m}}_{2.145\mu{\rm m}} f_\lambda d\lambda},
\end{equation}
where $f_\lambda$ is the spectrum's flux as a function of wavelength. The idealised spectroscopic CO index to first order differs from the photometric one only by an additive zeropoint \citep{1994MNRAS.269..655K}.

\subsection{Results} 
\label{sec:coindex:res}
\begin{figure}
\centering
\includegraphics[width=0.9\columnwidth]{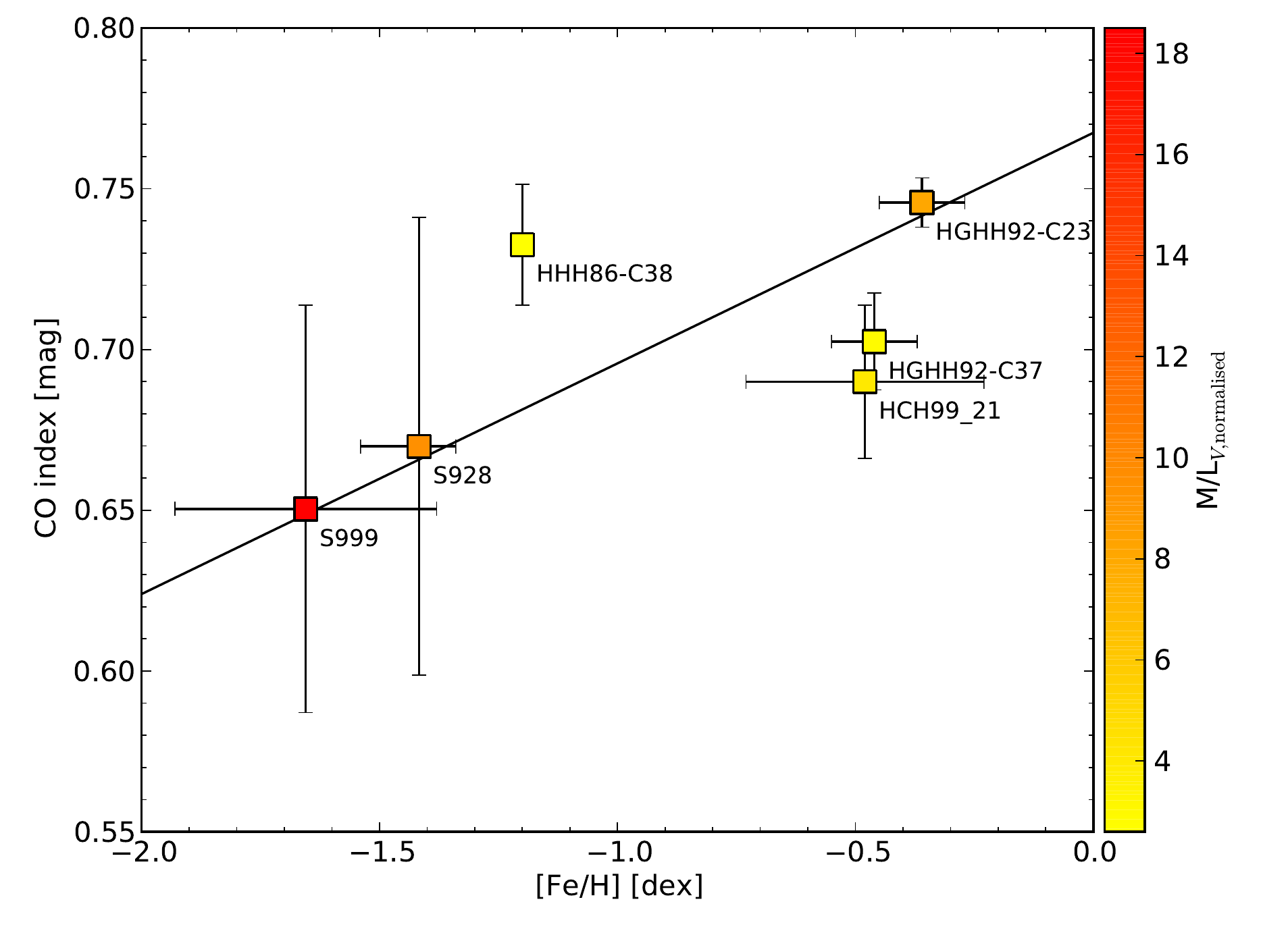}
\caption{The measured CO index measured versus the metallicity of two Virgo cluster UCDs (S999 and S928) and GCs/UCDs around NGC~5128. The colour scale refers to the normalised mass to light ratio of each object on the colour scale given on the right side (on which $\sim$4 is the maximum M/L$_V$ of a `normal' stellar population). The black line has the slope of the empirical relation for Galactic GCs between CO and [Fe/H] given by \citet{2001AJ....122.1896F}. Its location in the vertical direction (corresponding to the zeropoint between spectroscopic and photometric CO index) was fixed using a $\chi^2$-fit to all data. All data are within $\sim3\sigma$ of the empirical relation; in particular the high-M/L UCDs S999 and S928 do not lie significantly below the relation like it would be expected if the high M/L were due to a bottom-heavy stellar mass function.}
\label{coindex:COvsFeH_summation}
\end{figure}

Fig.~\ref{coindex:COvsFeH_summation} shows the measured CO index as a function of metallicity for each target. The adopted metallicities are based on either photometric colours or spectroscopy were assembled from the literature  \citep{2005ApJ...627..203H,2008ApJ...677..276M,1998ApJ...496..808C,2008MNRAS.386.1443B}. The colour scale reflects the objects' M/L$_{V}$, normalised to solar metallicity \citep[see][]{2008ApJ...677..276M}. On this scale, the maximum stellar population M/L$_{V}$ assuming a Kroupa or Chabrier IMF is $\sim4$.

The black line in Fig.~\ref{coindex:COvsFeH_summation} represents the empirical trend of the CO index with metallicity in Galactic GCs derived by \citet{2001AJ....122.1896F}. According to \citet{2008ApJ...677..276M}, if the high mass to light ratios of UCDs were due to an overabundance of low-mass stars, the strength of their CO absorption should deviate from this relation in the direction towards a lower CO index. For the highest M/L UCDs, such as the Virgo cluster UCDs S999 and S928, this offset should be $\ga0.05$\,mag.

The data show no such offset and all targets show a CO index compatible within $\la3\sigma$ with the empirical relation of Galactic GCs. The observations were originally designed to yield uncertainties in the CO index $<0.05$\,mag for the two Virgo cluster UCDs. However, as already mentioned, the modifications to the spectrum extraction that were necessary to suppress systematic biases in the low signal to noise regime, caused the uncertainties to be larger than the expected effect. 

In summary, the CO absorption in the high M/L Virgo cluster UCDs S999 and S928 is in agreement with that of the lower M/L comparison GCs/UCDs in NGC~5128 when the empirical dependence of the CO index on metallicity is taken into account. Thus, there is no evidence that the high M/L of the UCDs are caused by an overabundance of low-mass stars. Unfortunately, the uncertainties on the measured CO index of the high M/L UCDs are large compared to the expected deviations in the case of a bottom-heavy stellar mass function. Therefore, an overabundance of low-mass stars in these UCDs also cannot be firmly excluded.

\section{Extended globular clusters in the outer Galactic halo}
\label{sec:GalacticGCs}

\subsection{The Galactic GC system}
\label{sec:GCsystem}
The globular cluster system of the Milky Way extends out to more than 100\,kpc and contains more than 150 known GCs \citep[e.g.][2010 edition]{1996AJ....112.1487H}, with more GCs being discovered in current wide-field infra-red imaging surveys \citep[e.g.][]{2011A&A...527A..81M}. GCs have masses of $10^4\!\la\!\mathrm{M}\!\la\!2.5\times10^6\,\msun$. In terms of mass, GCs are not clearly delineated from the class of young or open clusters, such as Westerlund\,1 \citep[$\sim\!10^5\,\msun$;][]{2005A&A...434..949C}. Also the distinction between GCs and UCDs is somewhat arbitrary and the upper mass limit of $\sim\!2\times10^6\,\msun$ stated above is simply the mass of the most massive Galactic GC, $\omega$Cen \citep[NGC\,5139; the][]{2006A&A...445..513V}, and is generally adopted as the lower mass limit for UCDs for the same reason. 

As mentioned before, physical size alone is not sufficient to distinguish extended GCs from ultra-faint dwarf spheroidals (dSphs) and both classes overlap also in luminosity. If size and luminosity together are considered, the two classes stand apart in that GCs have higher surface brightnesses \citep[e.g.][]{2008MNRAS.390L..51V}. The most striking difference, of course, are the dynamical masses. Ultra-faint dSphs are thought to be the most dark matter dominated stellar systems known with $V$-band mass-to-light ratios of $10^2$ - $3\times10^3\mlsun$
\citep[and references therein]{2013pss5.book.1039W}. While also GCs were conjectured to have formed inside their own dark matter halos \citep{1984ApJ...277..470P}, current observations of their density profiles and velocity dispersions suggest that there is no detectable amount of dark matter in GCs \citep[e.g.][]{2009MNRAS.396.2051B,2011ApJ...741...72C,2011MNRAS.411.1536H,2012ApJ...751....6D}.

The Galactic GC system is not a homogeneous population of objects. A sub-population of metal-rich GCs ([Fe/H]$>-1.5$\,dex) is associated with the Galactic bulge or disc, but also metal-poor ($-2<$[Fe/H]$<-1.5$\,dex) GCs show an over-density in the bulge region \citep[e.g.][]{1999A&ARv...9..273V}. Also in the halo GC population several components can be distinguished. About one quarter of these belongs to the so-called `outer halo', at Galactocentric distances larger than 15\,kpc \citep[e.g.][]{2004MNRAS.354..713V}. Several of these outer halo clusters are also attributed to the `young halo' sub-population because they seem to be 1-2~Gyr younger than the old, inner halo GCs of similar metallicity \citep[e.g.][]{2010ApJ...708..698D}. A number of authors have suggested that the young and/or outer halo GCs were accreted by the Milky Way via the infall of dwarf satellite galaxies \citep[e.g.][]{1996ASPC...92..434M,2000ApJ...533..869C,2004MNRAS.355..504M,2007ApJ...661L..49L,2010MNRAS.404.1203F}, similar to the halo assembly scenario already proposed by \citet{1978ApJ...225..357S},  whereas the classical old GCs may have formed during an early and rapid dissipative collapse of the Galaxy's halo \citep{1962ApJ...136..748E}.

In recent years, an increasing number of GCs has been found to contain several stellar populations. In the majority of these clusters the different stellar subpopulations stand out by a difference in light element abundances, and younger populations are thought to have formed from material polluted by the first generation of GC stars \citep[see][for an overview]{2012A&ARv..20...50G}. However, in $\omega$Cen and M\,54 the sub-populations differ also significantly (by $\ga0.1$\,dex) in the iron abundance, and by several Gyr in age. Both clusters have been proposed to be the nuclei of accreted dwarf galaxies \citep[e.g.][]{1999Natur.402...55L,1995Obs...115..256B}. Therefore, these two clusters may be more similar to UCDs than to classical GCs. 

\subsubsection{Dynamical evolution of GCs}
\label{sec:GCsystem:evol} 
The dynamical evolution of GCs is driven by internal processes and influenced by the Galactic potential in which the clusters evolve. The early evolution of a star cluster is dominated by the mass-loss due to stellar winds and core-collapse supernovae with massive stellar progenitors, leading to an expansion of the cluster. Since lower-mass stars evolve more slowly, and since they retain a larger fraction of their initial masses, the importance of stellar evolution for the dynamical evolution of the clusters decreases with time. 

Two-body relaxation arises from encounters of stars, in which these exchange kinetic energy. Stars that lose kinetic energy sink into the cluster centre, while stars that gain kinetic energy populate orbits further away from the cluster's centre. The time-scale for two-body relaxation, at the half-mass radius, for almost all Galactic GCs is shorter than their respective ages \citep[e.g.][2010 edition]{1996AJ....112.1487H}. Thus, most Galactic GCs are mass segregated, i.e. more massive stars show a more concentrated radial distribution than lower-mass stars. As a consequence, the mass function of a cluster becomes shallower in the cluster centre, but also globally as stars at large radii are more easily lost to the Galactic potential \citep{1997MNRAS.289..898V,2003MNRAS.340..227B}. Comparing the overall slopes of the stellar mass functions observed in Galactic GCs with numerical simulations, \citet{2008ApJ...685..247B} found that several GCs are more strongly depleted in low-mass stars than expected from dynamical evolution if a \citet{2001MNRAS.322..231K} IMF is adopted. They argued that this could indicate primordial mass segregation in this clusters. Evidence for present-day mass segregation in two diffuse outer halo clusters with present-day half-mass relaxation times larger than the Hubble time, Palomar~4 and Palomar~14, and the question, whether this implies primordial mass segregation, is discussed in Section~\ref{sec:massseg}. 

\subsubsection{Remote GCs as probes for testing gravitational theories}
\label{sec:GCsystem:MOND} 
GCs have also been recognised as valuable probes for testing fundamental physics \citep[e.g.][]{2003A&A...405L..15S}. \citet{2005MNRAS.359L...1B} proposed to use diffuse outer halo GCs to distinguish between classical and modified Newtonian dynamics \citep[MOND;][]{1983ApJ...270..371M,1983ApJ...270..365M,1984ApJ...286....7B}. MOND is very successful in explaining the flat rotation curves of disk galaxies without any assumption of unseen dark matter, although it faces problems reproducing gravitational lensing from galaxy clusters and the anisotropies in the cosmic micro-wave background \citep{
2012LRR....15...10F}. According to MOND, Newtonian dynamics breaks down for accelerations lower than $a_0 \simeq 1\times10^{-8}$ cm\,s$^{-2}$ \citep{1991MNRAS.249..523B,2002ARA&A..40..263S}.

In remote Galactic GCs, the external acceleration due to the Galaxy experienced by the cluster stars is below this critical limit of $a_0$, and the radial velocity dispersion profiles of such clusters can thus be used to distinguish between MOND and Newtonian dynamics (see e.g. \citealt{2010MNRAS.401..131S}, who derived \citeauthor{1966AJ.....71...64K}-like models for GCs in MOND). It should be noted that \citet{2003A&A...405L..15S,2007A&A...462L...9S}, \citet{2010A&A...523A..43S} and \citet{2011A&A...525A.148S} reported a flattening of the velocity dispersion profile at accelerations comparable to $a_0$ also in GCs with Galactocentric distances $\la20$~kpc. Since the external acceleration in these clusters is well above $a_0$, such flattened velocity dispersion profiles in `inner' GCs, instead of being due to MONDian dynamics, might be due to tidal heating and unbound stars or to contamination by field stars \citep[e.g.][]{1998AJ....115..708D,2010MNRAS.401.2521L,2010MNRAS.406.2732L,2010MNRAS.407.2241K}.

Two GCs in the outer halo that have been studied extensively in the context of testing MOND are the massive cluster NGC~2419 and the diffuse cluster Pal\,14. 

Based on radial velocities of 40 member stars of NGC~2419 \citet{2009MNRAS.396.2051B} derived a dynamical mass of 9$\pm2\times10^{5}\msun$, compatible with the photometric expectation from a simple stellar population with a \citet{2001MNRAS.322..231K} IMF. Moreover, they found no flattening of the velocity dispersion profile at low accelerations that could point to MONDian dynamics or dark matter in this cluster. \citet{2011ApJ...738..186I} studied an extended radial velocity sample of 178 stars of NGC~2419 and found that, while radial anisotropy is required in both Newtonian and MONDian dynamics to explain the observed kinematics, the data favour Newtonian dynamics, with their best-fitting MONDian model being less likely by a factor of $\sim40,000$ than their best-fitting Newtonian model. \citet{2012MNRAS.419L...6S} challenged this conclusion, arguing that in MONDian dynamics non-isothermal models, approximated by high-order polytropic spheres, can reproduce the cluster's surface brightness and velocity dispersion profiles. This led \citet{2011ApJ...743...43I} to extend the analysis of their data to polytropic models in MOND. Again, they concluded that the best-fitting MONDian model is less likely by a factor of $\sim5000$ than the best-fitting Newtonian model, and that the data therefore pose a challenge to MOND, unless systematics are present in the data \citep[but see also][]{2012MNRAS.422L..21S}. 

In the most diffuse outer halo clusters, i.e. clusters with large effective radii, low masses and therefore low stellar densities, also the internal acceleration due to the cluster stars themselves is below $a_0$ throughout the cluster. In these clusters, instead of only the shape of the velocity dispersion profile, also the global velocity dispersion can be used to discriminate between MOND and Newtonian dynamics. The expected global velocity dispersions differ by up to a factor of 3 between both models
\citet{2005MNRAS.359L...1B,2011A&A...527A..33H}. For the diffuse halo cluster Pal\,14, \citet{2009AJ....137.4586J} measured radial velocities of 16 cluster members stars and found a good agreement between the cluster's photometric and dynamical mass in classical Newtonian dynamics, while the velocity dispersion is significantly lower than predicted in MOND \citep{2009MNRAS.395.1549H}. \citet{2010A&A...509A..97G} challenged this conclusion and argued on the basis of a Kolmogorov-Smirnov (KS) test that the sample of radial velocities (or, alternatively, the sample of studied diffuse outer halo GCs) is too small to rule out MOND

Measurements of the dynamical and photometric mass of the diffuse outer halo cluster Pal\,4 \citep{2012MNRAS.423.2917F} in the context of testing MOND will be summarised in Section~\ref{sec:pal4} 

\subsection{Mass-segregation in Palomar~4 and Palomar~14}
\label{sec:massseg}
As mentioned above, several Galactic GCs appear to be more dynamically more evolved than expected from their present half-mass relaxation times \citet{2008ApJ...685..247B}. This motivated a study to clarify the presence of mass segregation in the two extended outer halo clusters Pal\,4 \citep{2012MNRAS.423.2917F} and Pal\,14 \citep{Pal14prep}. 

With a Galactocentric distance of 103\,kpc Pal\,4 is the second to outermost halo GC after AM\,1 \citep[at 123\,kpc;][2010 edition]{1996AJ....112.1487H}. With a half-light radius of 18\,pc, Pal\,4 also among the most extended Galactic GCs. Pal\,14 is somewhat closer to the Galactic centre ($\sim$66\,kpc), but with a half-light radius of 46\,pc, is even more extended \citep{2011ApJ...726...47S}. Both cluster's sizes and luminosities are comparable to some of the Galaxy's ultra-faint dSph satellites. With a metallicity of [Fe/H]$=-1.41\pm0.17$\,dex \citep{2010A&A...517A..59K}, and [Fe/H]=$-1.3$\,dex \citep{2012A&A...537A..83C}, and an age of $11 \pm 1$~Gyr \citep{2012MNRAS.423.2917F} and 11.5$\pm0.5$\,Gyr \citep{2009AJ....137.4586J}, respectively, both clusters can be considered part of `young halo' population sub-population. 

Due to their diffuse structure, their present-day half-mass relaxation times are $\sim$14~Gyr (Pal\,4) and $\sim$20~Gyr (Pal\,14), respectively. Thus, one would naively expect that two-body relaxation has not affected these clusters significantly, yet. In agreement with this expectation, \citet{2011ApJ...737L...3B} found that the distribution of blue straggler stars in Pal\,14 is not centrally concentrated with respect to less massive HB and RGB stars in this cluster. On the other hand, \citep{2009AJ....137.4586J} found that the mass function of main-sequence stars in the inner part of Pal\,14, has a power-law slope of $\alpha=1.3\pm0.4$, and thus is strongly depleted in low-mass stars compared to a Kroupa IMF. Performing direct $N$-Body simulations of Pal\,14, \citet{2011MNRAS.411.1989Z} found that the effect from two-body relaxation in such an extended cluster is too weak to reproduce this shallow present-day mass function slope in models starting with a Kroupa IMF. They also showed that observed mass-function of low-mass stars can be reproduced by models with a high degree of primordial mass segregation. In this case, the cluster should also show mass segregation at the present.

\subsubsection{Observations and analysis}
\label{sec:massseg:data}
To answer the question, whether clusters like Pal\,14 and Pal\,4 are mass-segregated, photometry was obtained from archival HST imaging taken in programs 5672 \citep[Pal 4; cf.][]{1999AJ....117..247S} and 6512 \citep[Pal\,14; cf.][]{2008AJ....136.1407D}. Both datasets consist of deep F555W ($V$) and F814W ($I$) Wide Field Planetary Camera 2 (WFPC2) exposures. PSF-fitting photometry, along with artificial star tests in order to estimate completeness and photometric uncertainties, was obtained using the \textsc{HSTPHOT} package \citep{2000PASP..112.1383D}. 

The stellar mass function as a function of radius in each cluster was obtained by converting stellar magnitudes to masses using a \citet{2008ApJS..178...89D} isochrone with the corresponding chemical composition and age, shifted to the distance and reddening of each cluster. Completeness corrections were applied for photometric incompleteness and geometric coverage of the WFPC2 data. The stellar mass range from the tip of the red giant branch to the $\sim50$\,per cent completeness limit at the faint end (${\rm F555W}\la27.6{\rm\,mag}$ in Pal\,4, ${\rm F555W}\la27.1{\rm\,mag}$ in Pal\,14), corresponds to $0.55\le m/\msun\le 0.85$ in both clusters. In this mass range, the stellar mass function at different radii of both clusters is represented by a single power law. 

\subsubsection{Results and discussion}
\begin{figure}
\centering
\includegraphics[width=0.9\columnwidth]{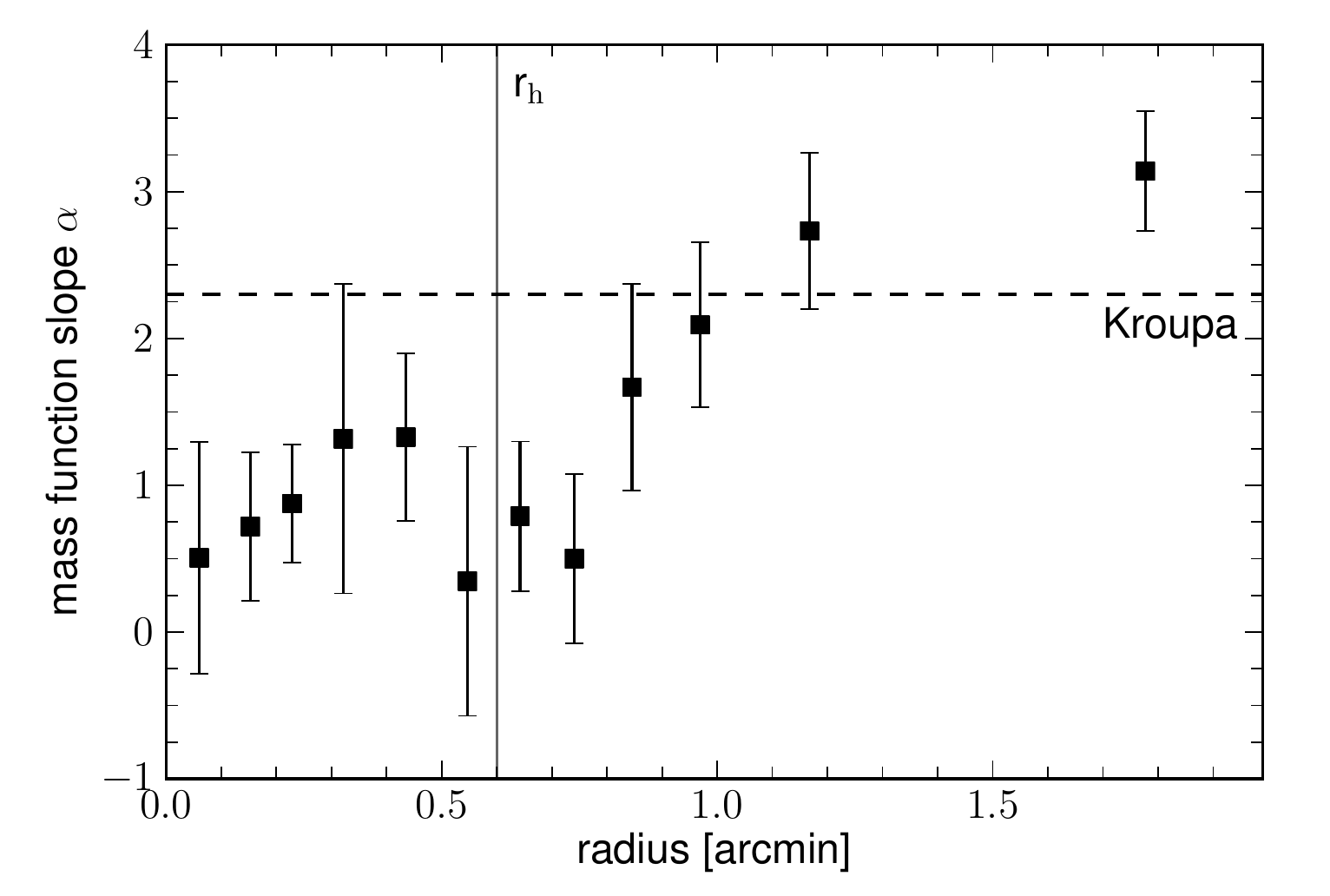}
\caption{Mass segregation in Pal\,4. The data show the best-fitting power-law mass function slope as a function of radius \citep{2012MNRAS.423.2917F}. There is a clear trend of the mass function to steepen with increasing radius. In the central $<0.8$\,arcmin, the mass function is significantly depleted in low-mass stars compared to a Kroupa mass function ($\alpha=2.3$ in this mass range; indicated by the dashed line). Beyond $\sim$1.2\,arcmin, the mass function is steeper than that.}
\label{pal4:masssegregation}
\end{figure}

\begin{figure}
\centering
\includegraphics[width=0.9\columnwidth]{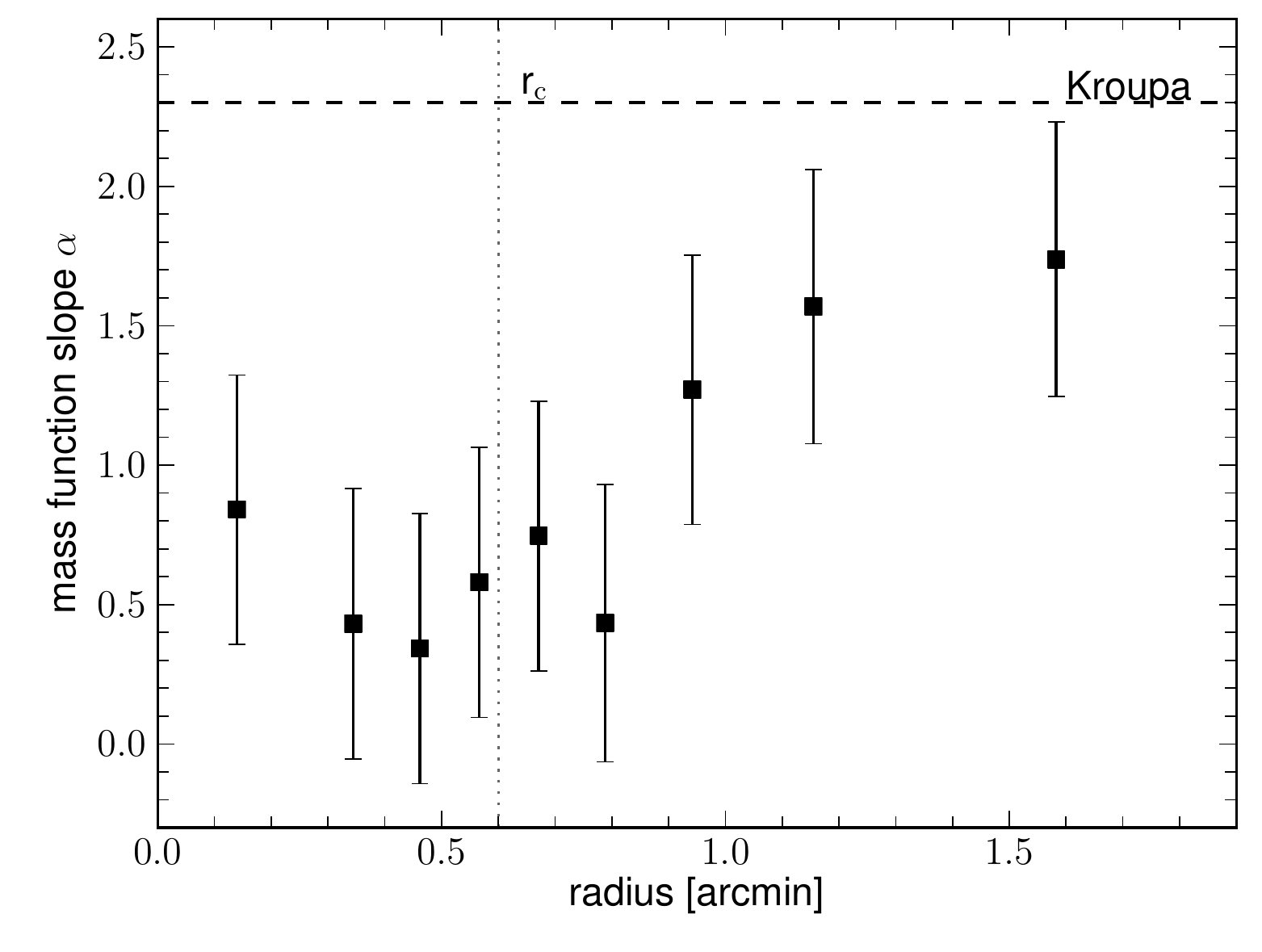}
\caption{Same as Fig.~\ref{pal4:masssegregation}, but for Pal\,14. The dotted vertical line indicates the cluster's \emph{core} radius (0.6\,arcmin); its half-light radius (2.2\,arcmin) is not covered by the WFPC2 data.}
\label{pal14:masssegregation}
\end{figure}

The best-fitting power-law slope of the mass function in Pal\,4, as a function of radius is shown in Fig.~\ref{pal4:masssegregation}. Fig.~\ref{pal14:masssegregation} shows the same for Pal\,14. The angular radial coverage of the WFPC2 data is similar in both clusters. While in the case of Pal\,4, the data contain the cluster's half-light radius of 0.6\,arcmin (18\,pc), in the case of Pal\,14, the half-light radius (2.2\,arcmin or 46\,pc) lies beyond the footprint of the data. 

In both cases, the mass function steepens with distance from the cluster centre, indicating that mass segregation indeed is present in Pal\,4 and Pal\,14. The present-day half-mass relaxation times of both clusters are larger than their respective ages. Moreover, specifically for Pal\,14, direct $N$-body simulations showed that two-body relaxation alone is insufficient in order to deplete the mass function of low-mass stars if the cluster was born with no initial mass-segregation and an initial half-mass radius similar to its present-day value \citep{2011MNRAS.411.1989Z}. This may suggest that primordial mass segregation occurs also in globular clusters, similar to what is observed in several young Galactic clusters \citep[e.g.][]{1988MNRAS.234..831S,1997AJ....113.1733H,2011MNRAS.413.2345H}. 

Alternatively, it is also possible that the mass segregation in Pal\,14 and Pal\,4 is due to two-body relaxation, if the clusters had significantly shorter relaxation times in the past. This would be satisfied of the clusters were more compact (by a factor of $\sim\!2$) for most of their evolution. In this case, the clusters may have been inflated to their present-day size as a result of tidal shocks on their orbit through the Galactic potential \citep{1999ApJ...522..935G}, similar what has been proposed for the currently dissolving cluster Pal\,5 \citep{2004AJ....127.2753D}. The latter has a present Galactocentric distance of $18.5$\,kpc and is believed to be near its apogalacticon \citep{2003AJ....126.2385O}. In order to be similarly affected by disk or bulge shocks, the presently much more remote Pal\,4 (103\,kpc) and Pal\,14 (66\,kpc) would have to be on highly eccentric orbits. That Pal\,14 is affected by tides, and losing stars even at its current remote location is evidenced by the tidal tails discovered around the cluster \citep[][]{2010A&A...522A..71J,2011ApJ...726...47S}. Observational constraints on the orbits of Pal\,14 and Pal\,4, which may be feasible in the GAIA \citep{2001A&A...369..339P} era, will help clarify the question, whether primordial mass segregation occurs in GCs, or whether also these extended clusters once had a more compact structure, like it is typical for most GCs.  

\subsection{The dynamical and photometric mass of Palomar~4}
\label{sec:pal4}
Pal\,4 was also proposed as a good test case for MONDian dynamics \citep{2005MNRAS.359L...1B,2011A&A...527A..33H}. Due to its diffuse structure, the internal acceleration due to the cluster stars themselves is below $a_0$ throughout the cluster. Therefore, not only the shape of the velocity dispersion profile (like in the case of NGC~2419), but also the \emph{global} velocity dispersion of Pal\,4 \citep[just like that of Pal\,14;][]{2009AJ....137.4586J}, can be used to discriminate between MONDian and Newtonian dynamics. 

\subsubsection{Observations and analysis}
In order to measure the cluster's velocity dispersion, radial velocities of 24 candidate cluster members (19 likely red giants, 5 potential asymptotic giant branch stars) in the direction of Pal\,4 were obtained using the Keck High Resolution Echelle Spectrograph \citep{1994SPIE.2198..362V}. The spectra covered the wavelength range from 445 to 688\,nm at a resolution of $R=45000$. Exposure times varied between 300 and 2400\,s, depending on the target magnitude ($17.8<V<19.9{\rm\,mag}$). The targets were selected from pre-images obtained with the Keck Low-Resolution Imaging Spectrometer. Both datasets were part of a larger program aiming to study the internal kinematics of outer halo GCs \citep[see][]{2002ApJ...574..783C}. Radial velocities of the candidate cluster stars were measured via cross-correlation \citep{1979AJ.....84.1511T}. Based on a clean sample of individual radial velocities, the mean heliocentric radial velocity and velocity dispersion of Pal\,4 were derived to be $v_r=72.55\pm0.22\kms$ and $\sigma=0.87\pm0.18\kms$, respectively. 

The cluster's photometric mass was estimated by extrapolating the stellar masses derived from the HST photometry in Section~\ref{sec:massseg}. Summing stellar masses in the mass range $0.55\le m/\msun\le 0.85$ yielded a mass of $5960\pm110\,\msun$ contained inside the radius covered by the WFPC2 pointing, $r<2.26$\,arcmin. This mass was then extrapolated to lower-mass stars, assuming that the observed mass-function slope ($\alpha=1.40 \pm 0.25$ in the WFPC2 pointing) holds down to 0.5~$\msun$ and adopting a \citet{2001MNRAS.322..231K} mass function for masses $0.01\le
m/\msun\le 0.5$. This yielded an the extrapolated stellar mass in the mass range $0.01\le m/\msun\le 0.85$ is $14500\pm1300\,\msun$. For higher mass stars, assuming that stars with initial masses $0.85\le m \le
8\msun$ have formed 0.6\,$\msun$ white dwarfs, a similar extrapolation yields an additional mass of $\mathrm{M}_\mathrm{WD}=8900\pm800\,\msun$ in white dwarfs. 

Based on fitting a \citet{1966AJ.....71...64K} density profile to the LRIS and HST photometry, it was estimated that  $98.3\pm0.4$ per cent of the cluster's light lies within $r=2.26$\,arcmin, the maximum radius covered by the WFPC2 data. Approximating further that mass follows light, and extrapolating the observed mass out to to the tidal radius, the total photometric mass of Pal\,4 amounts to $\mathrm{M}_\mathrm{phot}=29800\pm800\,\msun$, including the corrections for low-mass stars and white dwarfs.

\subsubsection{Results and discussion}
Using $N$-body simulations \citet{2011A&A...527A..33H} calculated the dynamical mass of Pal\,4 as a function of the global line-of-sight velocity dispersion in both Newtonian and MONDian dynamics. For the measured line-of-sight velocity dispersion of $0.87\pm0.18\kms$, the theoretically predicted mass in
MOND is $\mathrm{M}_\mathrm{MOND}=3900^{+1400}_{-1500}\,\msun$ and in Newtonian dynamics
$\mathrm{M}_\mathrm{Newton}=32000\pm\,13000\msun$.  

Compared to the measured photometric mass, $\mathrm{M}_\mathrm{phot}=29800\pm800\,\msun$, the Newtonian dynamical mass matches very well, while the MONDian prediction is significantly too low. The Newtonian case is therefore favoured by the observational data. On the other hand, a Kolmogorov-Smirnov test comparing the observed distribution of radial velocities with the predicted line-of-sight velocity distribution in Newtonian and MONDian dynamics \citep[following the argument of][to take into account the stochasticity of the small radial velocity sample]{2010A&A...509A..97G}, yields a probability of 87\,\% and 19\,\% for the data to be drawn from the Newtonian and MONDian distribution, respectively. Thus, while favouring Newtonian dynamics, the current observational data on Pal\,4 alone do not rule out MOND, either. Nevertheless, they add to the body of evidence (after Pal\,14 and NGC\,2419) that the outer halo clusters obey Newtonian dynamics rather than MOND. 

Given the excellent match between photometric and (Newtonian) dynamical masses, there is no need to invoke the presence of dark matter in Pal\,4,
although a small amount of dark matter cannot be excluded. And indeed, Pal\,4 has a `classical' star cluster mass to light ratio of $\mathrm{M}_\mathrm{Newton}/\mathrm{L}_{V}$ of $\sim1.6\,\msun\,\lsun^{-1}$, despite its structural similarity to the dark matter dominated ultra-faint dSphs.

\section{Summary \& Outlook}
\label{sec:summary}
The current literature on ultra-compact dwarf galaxies (Section~\ref{sec:UCDsproperties}) suggests that they are a mix of massive star clusters and of (remnants of) tidally stripped galaxies. In either case, the origin of the higher than expected dynamical masses of massive UCDs remains unclear. If these objects are remnant nuclei, the presence of massive black holes appears to be a promising solution \citep{2013A&A...558A..14M}, which can be tested by measuring their spatially resolved kinematics. The results on UCD3 (Section~\ref{sec:ucd3}) demonstrate the feasibility of obtaining such data. While the studied UCD3 has a `normal' mass to light ratio, and accordingly, the data constrain its maximum black hole mass to be $\la5$\,per cent, similar observations of high-M/L UCDs will help test the black hole scenario. A search for evidence for a bottom-heavy stellar mass function in two high-M/L UCDs (Section~\ref{sec:coindex}) yielded a non-detection, although more data would be needed to rule out such a scenario.

At another interface between the regimes of star clusters and galaxies, the extended outer halo GC Pal\,4 behaves less surprisingly (Section~\ref{sec:pal4}): its dynamical mass, assuming Newtonian dynamics, is in excellent agreement with its photometric mass. Moreover, the data favour Newtonian dynamics over MOND, although they do not significantly rule out MOND. The finding of mass segregation in Pal\,4 and Pal\,14 (Section~\ref{sec:massseg}) could mean that primordial mass segregation is a common pattern in diffuse GCs. Alternatively, the present-day half-mass relaxation times of these clusters may not be representative because the clusters were more compact in the past. To answer the question, whether the clusters could have expanded due to disk or tidal shocks, information on their orbits will be required.

\acknowledgements
I thank the committee of German Astronomical Society (AG) for the Doctoral Thesis Award 2013. I would like to thank my advisers, Eva Grebel, Michael Hilker and Steffen Mieske for their continuous support. Holger Baumgardt, Andreas K\"upper, Patrick C\^ot\'e, Hosein Haghi, Pavel Kroupa, S.\,G.\,Djorgovski, and Leopoldo Infante are thanked for their contributions to part of the research presented here. I gratefully acknowledge support from the German Research Foundation (DFG) via Emmy Noether Grant Ko 4161/1.  

Based in part on observations made with the NASA/ESA Hubble Space Telescope, obtained
from the Multimission Archive at the Space Telescope Science Institute (MAST). STScI is operated by the Association of Universities for Research in Astronomy, Inc., under NASA contract NAS5-26555.  Some of the data presented herein were obtained at the W.M. Keck Observatory,
which is operated as a scientific partnership among the California Institute
of Technology, the University of California and NASA. The Observatory was made
possible by the generous financial support of the W.M. Keck Foundation.

\newpage

\end{document}